\documentclass[epj]{svjour}
\usepackage{epsf}
\begin{document}

\title{Coulomb repulsion versus Hubbard repulsion in a disordered chain}
\titlerunning{Coulomb Repulsion in one dimension}

\author{Franck Selva\thanks{\email{selva@drecam.saclay.cea.fr}}
        \and Jean-Louis Pichard}
\authorrunning{Franck Selva et al.}
        
\institute{CEA, Service de Physique de l'\'Etat Condens\'e, 
Centre d'\'Etudes de Saclay, 91191, Gif sur Yvette cedex, France}

\date{march 20, 2000}

\abstract{
We study the difference between on site Hubbard and long range 
 Coulomb repulsions for two interacting particles in a disordered chain. 
 While Hubbard repulsion can only yield weak critical chaos with 
 intermediate spectral statistics, Coulomb repulsion can drive the two 
 particle system to quantum chaos with Wigner-Dyson spectral statistics. 
 For intermediate strengths $U$ of the two repulsions in one dimension, 
 there is a crossover regime where delocalization and spectral rigidity 
 are maximum, whereas the limits of weak and strong $U$ are characterized 
 by a stronger localization and uncorrelated energy levels.
}

\maketitle

\section{Introduction}
\label{sec:intro}
  
 In low dimensions ($d \leq 2$) disorder always yields \cite{aalr} 
a finite localization length $L_1$ when the particles do not interact 
and there is no spin-orbit scattering. When one wants to study 
the role of electron electron interaction, a first issue is to know 
what kind of interaction is appropriate. When there are many 
carriers inside a large length $L_1$ (large density and weak 
disorder), it looks reasonable to assume weakly interacting Landau 
quasi-particles and to take the usual short range screened Coulomb 
repulsion. But, for low carrier densities ($10^{10}$-$10^{11}$ carriers 
per $cm^2$ is nowadays achieved \cite{kravchenko} in two dimensional 
heterostructures) the screening of the charges is somewhat 
problematic and one may find safer to consider bare long range Coulomb 
repulsion. One has in this case a system having charge crystallization 
as a natural limit when kinetic energy becomes negligible compared 
to Coulomb energy. The range of the interaction can also be varied 
by metallic gates located in the vicinity of the electron gas, as it 
is often done for having a tunable carrier density. This gives us the 
motivation to study the difference between on site Hubbard like repulsion 
and long range Coulomb repulsion in a simple limit: two electrons 
in a disordered chain. Since on-site interaction plays a role only if 
the orbital part of the wave function is symmetric, we restrict our study 
to the case with opposite spins.  

  The problem of two interacting particles (TIP) in one dimension 
has been mainly studied with on site Hubbard repulsion of strength $U$. 
As proposed by Shepelyansky \cite{shepelyansky} it has been  
numerically proven \cite{moriond,wmgpf,Voppen,Frahm} that interaction 
delocalizes a 
certain number of TIP states over a length $L_2 >> L_1$. For a ``contact'' 
interaction as Hubbard repulsion, the TIP system exhibits remarkable 
properties \cite{tip1,tip2,tip3,tip4}. The mixing of the one body states 
inside a scale $L=L_1$ and the associated delocalization effect for sizes 
$L \geq L_1$ is maximum for $U \approx U_c$, where $U_c$ is 
the fixed point of a duality transformation \cite{tip2} mapping 
the weak $U/t$ limit onto the weak $t/U$ limit, $t$ being the kinetic 
energy scale. For the one body problem, 
the spectral statistics contains important information: 
an Anderson insulator has uncorrelated levels (Poisson statistics) whereas a 
disordered metal displays Wigner-Dyson rigidity characterizing 
quantum chaos. At the mobility edge, lies a scale invariant critical 
statistics \cite{shapiro}, which exhibits a weaker spectral rigidity 
associated to weak critical chaos. For the TIP system with Hubbard 
interaction, the spectrum is Poissonian when $U \rightarrow 0$ and $U 
\rightarrow \infty$ 
and becomes \cite{tip2} more rigid when $U \approx U_c$. However, the maximum 
possible rigidity does not correspond to Wigner-Dyson rigidity, but 
to an intermediate rigidity analogous to those characterizing the one body 
spectrum at a mobility edge (critical statistics). It was noticed in 
Ref. \cite{tip3} that those intermediate statistics are also related to 
very slow interaction induced TIP diffusion at scales $L_1 \geq L \geq L_2$. 
We show in this study that, 
in contrast to Hubbard repulsion, Coulomb repulsion can drive the 
TIP one dimensional system to full quantum ergodicity with Wigner-Dyson 
statistics. From a statistical study of the interaction matrix elements 
coupling two free particle (2FP) states (i.e. the TIP eigenstates 
at $U=0$), one finds that Coulomb repulsion mainly favors hopping terms 
between 2FP states nearby in energy, with energy separation of the order 
of the TIP level spacing $\Delta_2 \propto L^{-2}$. Hubbard repulsion 
mainly induces \cite{tip2} hopping terms between 2FP states 
separated by a larger energy $\Delta_2^{eff} > \Delta_2$, and the measure 
of the coupled 2FP states is multifractal \cite{tip1}. 
However, the generic behavior of a TIP system with either Hubbard or Coulomb 
repulsions can be summarized by three regimes: a free particle limit 
dominated by 
Anderson localization; a large interaction limit dominated again by 
Anderson localization for Hubbard and by charge crystallization for 
Coulomb;and between those two Poissonian limits lies an intermediate regime 
characterized by 
a maximum mixing of the one particle states and a maximum delocalization 
effect. The intermediate regime in both cases is located around 
the interaction strengths $U_c$ for which the TIP system has participation 
ratios of same order in both preferential eigenbases characterizing the 
weak and strong interaction limits. 

\section{TIP Hamiltonian}
\label{sec:Hamiltonian}

 The TIP Hamiltonian ${\cal H}$ is given by the sum of two terms: 
the first ${\cal H_0}$ gives the kinetic energy (parameter $t$) and the random 
potentials (parameter $W$) in which the two particles can move, 
\begin{equation}
{\cal H_0}= -t \sum_{\{i,j\}} c^{+}_{i}c^{\phantom{+}}_{j}+ W \sum_i v_i n_i. 
\end{equation}
$v_i$ is randomly taken in the interval $[-\frac{1}{2},\frac{1}{2}]$, 
$c^{+}_{i}$ creates a particle on the site i and 
$n_i=c^{+}_{i}c_{i}$.
The second term ${\cal U}$ is the two body repulsion, which 
can be either on site Hubbard repulsion: 
\begin{equation}
{\cal U}= U \sum_i n_i(n_i-1)
\end{equation}  
or long range Coulomb repulsion:
\begin{equation} 
{\cal U} =  U \sum_i n_i(n_i-1) + \frac{U}{2} 
\sum_{\stackrel{i,j=1}{\rm |i-j|\le L/2}}^L
\frac{n_i n_j}{|i-j|} 
\end{equation} 
 The convention in this work is that two particles at 
the same site cost an energy $2U$ (and not $U$ as assumed in previous 
refs. \cite{tip1,tip2,tip3,tip4}). An additional cost of energy $U/p$ 
has to be paid by two particles separated by a distance $p$ with $L/2 
\geq p \geq 1$ when there is Coulomb repulsion. The boundary conditions 
(BCs) are taken periodic.

Let us denote ($\epsilon_{\alpha}, \psi_{\alpha}$) and 
$(E_{\alpha\beta},\psi_{\alpha\beta})$ the eigenenergies and  
eigenfunctions of the one particle state $|\alpha\rangle$ and 
of the 2FP state 
$|\alpha\rangle\bigotimes|\beta\rangle=|\alpha\beta\rangle$ 
respectively. One has $E_{\alpha\beta}= \epsilon_{\alpha}+\epsilon_{\beta}$.
The 2FP level spacing is $\Delta_2 \approx 2B/L(L+1)$ where the 
band width $B \approx 8t+2W$. The one particle localization length $L_1$ 
is defined from the weak disorder formula $L_1=100/W^2$. Hereafter, the 
energies will be given in units of the kinetic energy hopping term $t$ 
restricted to nearest neighbors. 

\section{TIP density of states}
\label{sec:DOS}

\begin{figure}
\vspace{-0.5cm}
\centerline{ 
\epsfxsize=4.2cm \epsfysize=4.2cm 
\epsffile{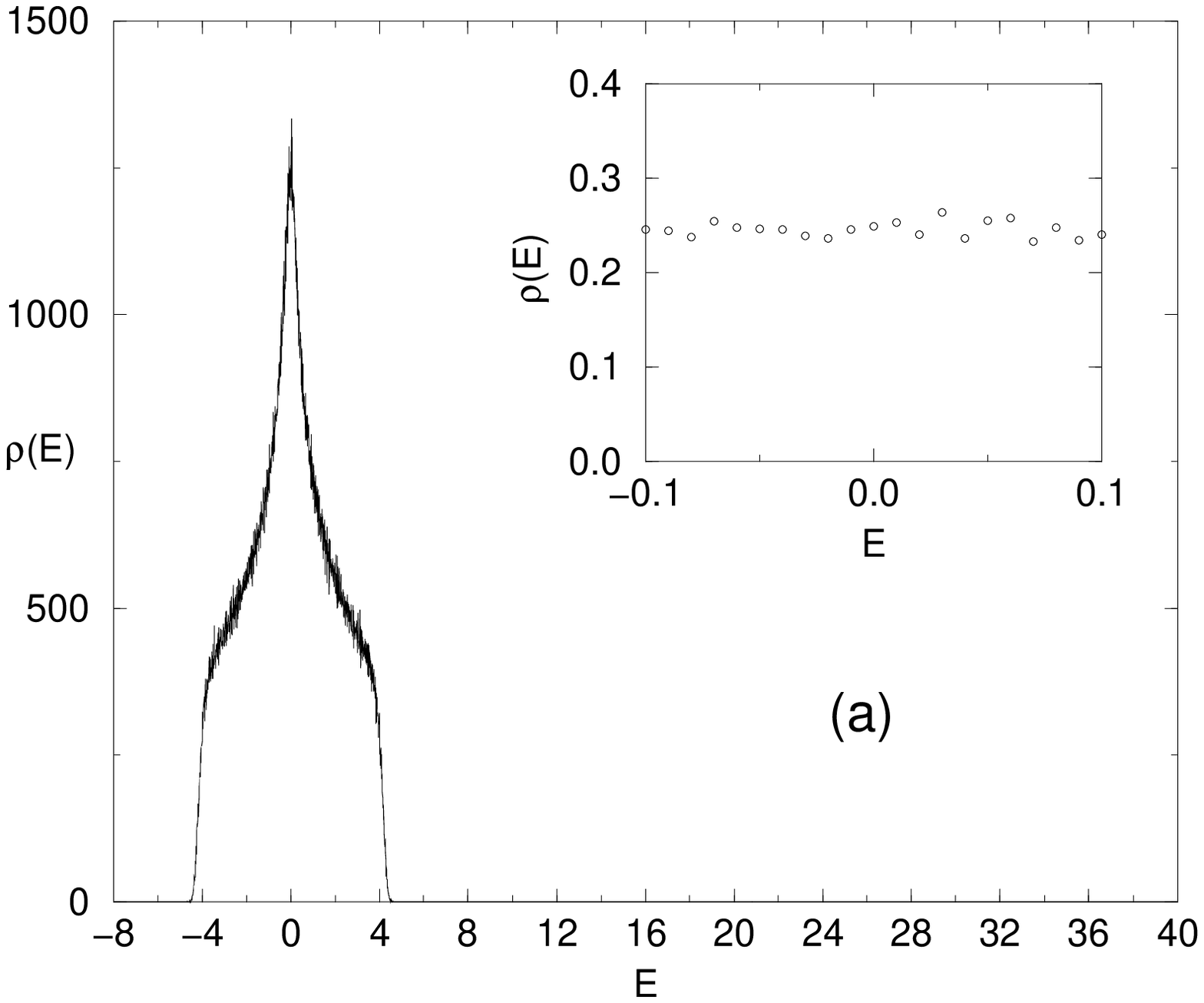}
\hfill 
\epsfxsize=4.2cm  \epsfysize=4.2cm 
\epsffile{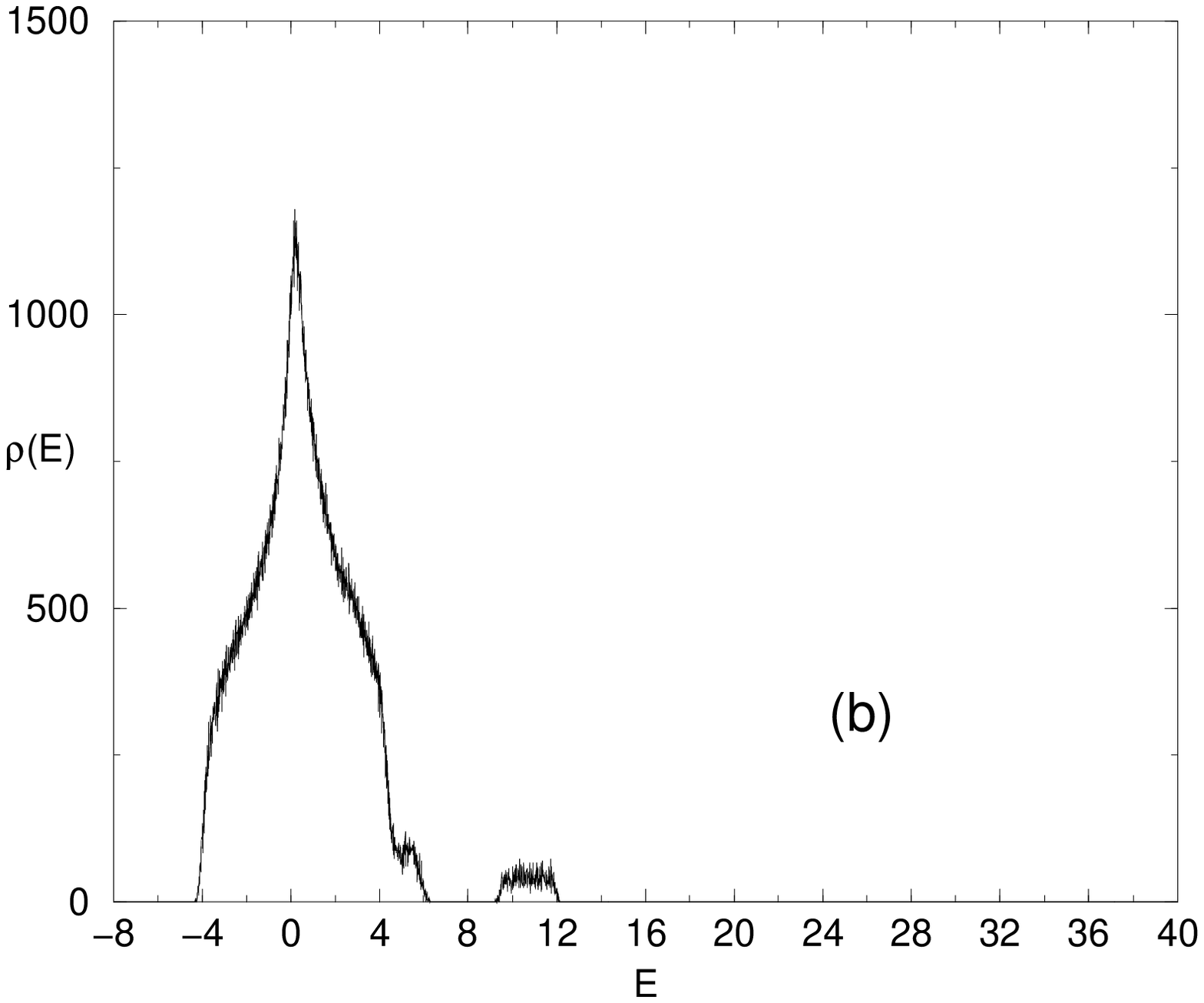}}
\centerline{
\epsfxsize=4.2cm  \epsfysize=4.2cm 
\epsffile{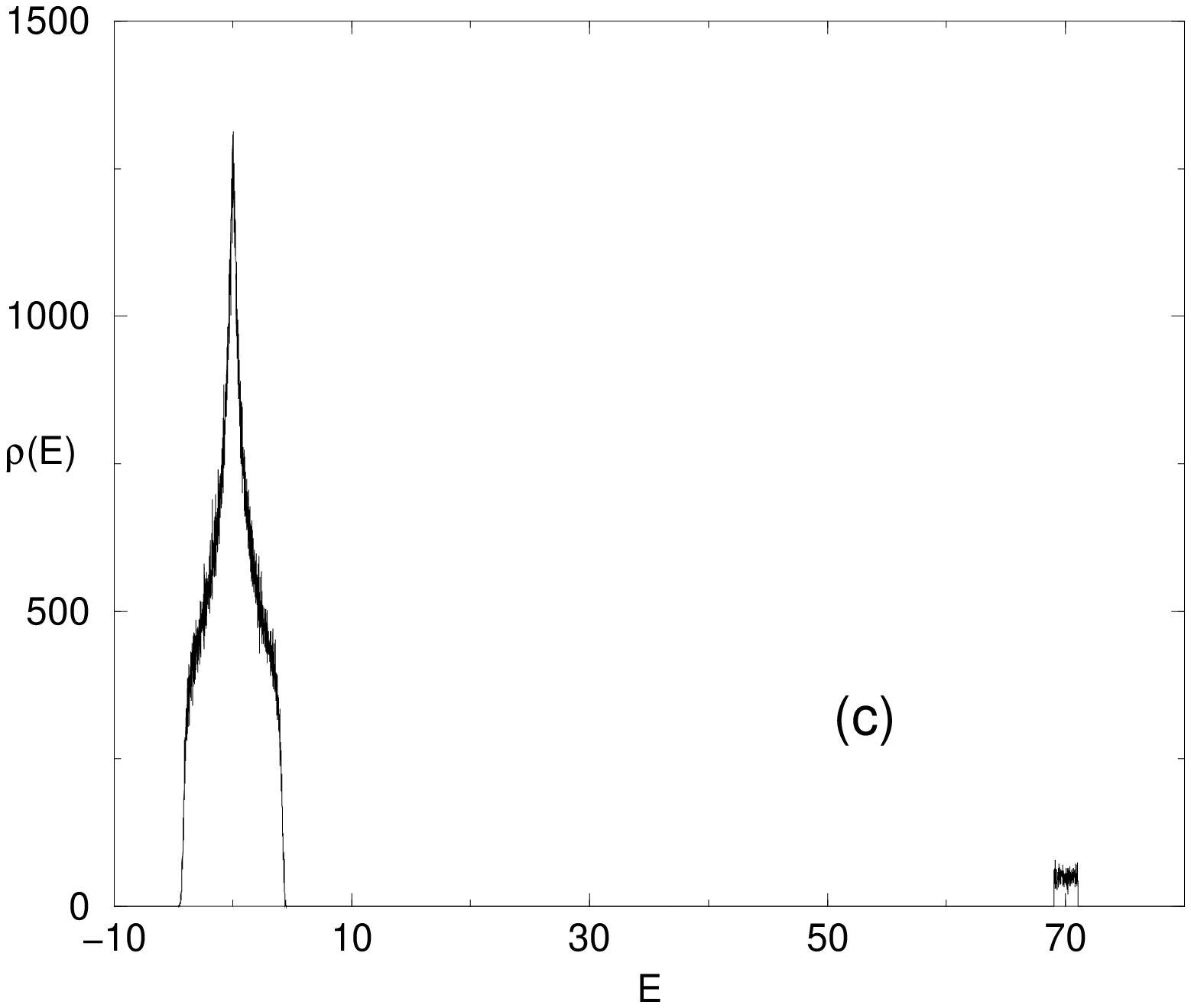}
\hfill 
\epsfxsize=4.2cm  \epsfysize=4.2cm 
\epsffile{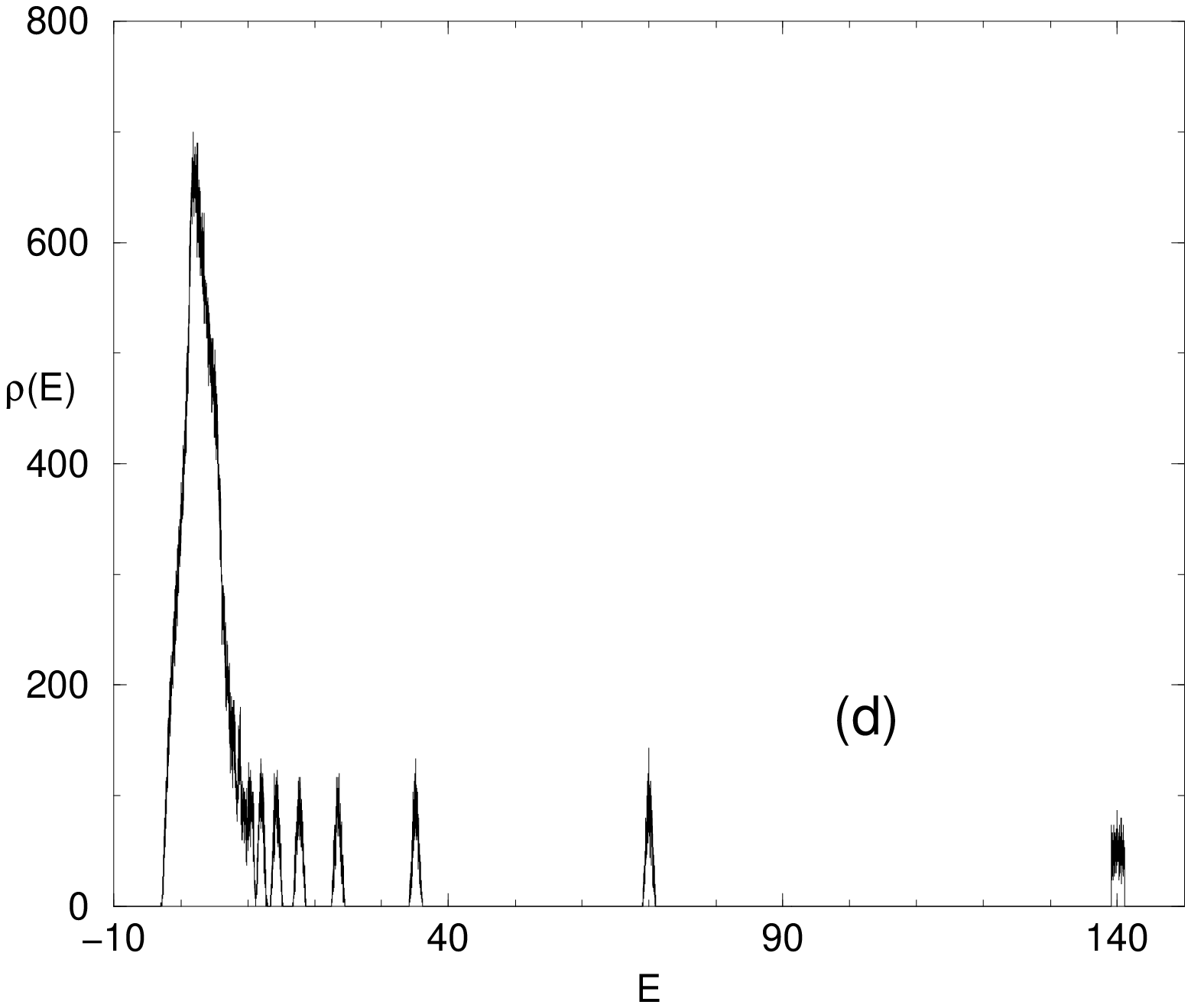}} 
\centerline{
\epsfxsize=4.2cm  \epsfysize=4.2cm 
\epsffile{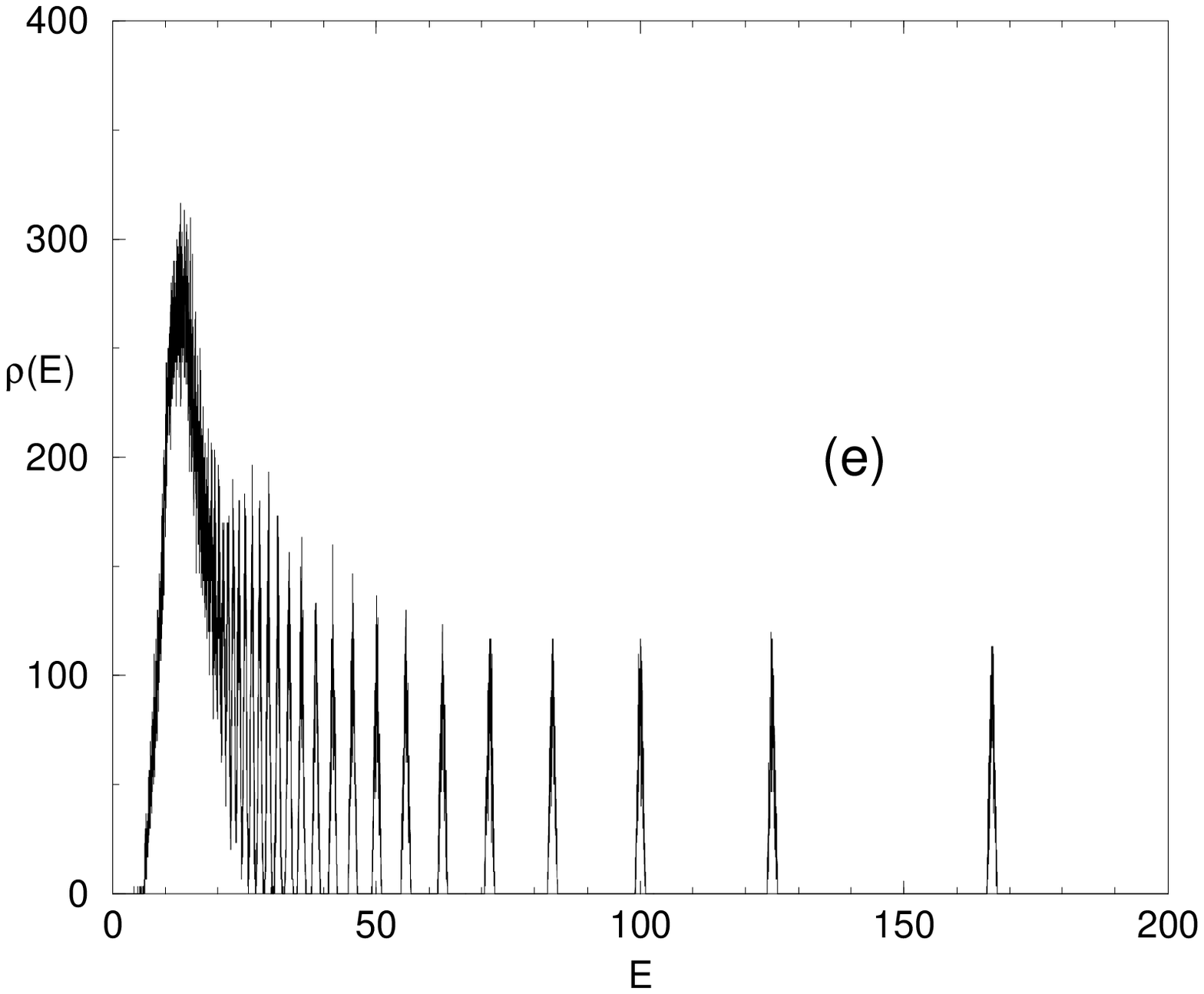}
\hfill 
\epsfxsize=4.3cm  \epsfysize=4.3cm 
\epsffile{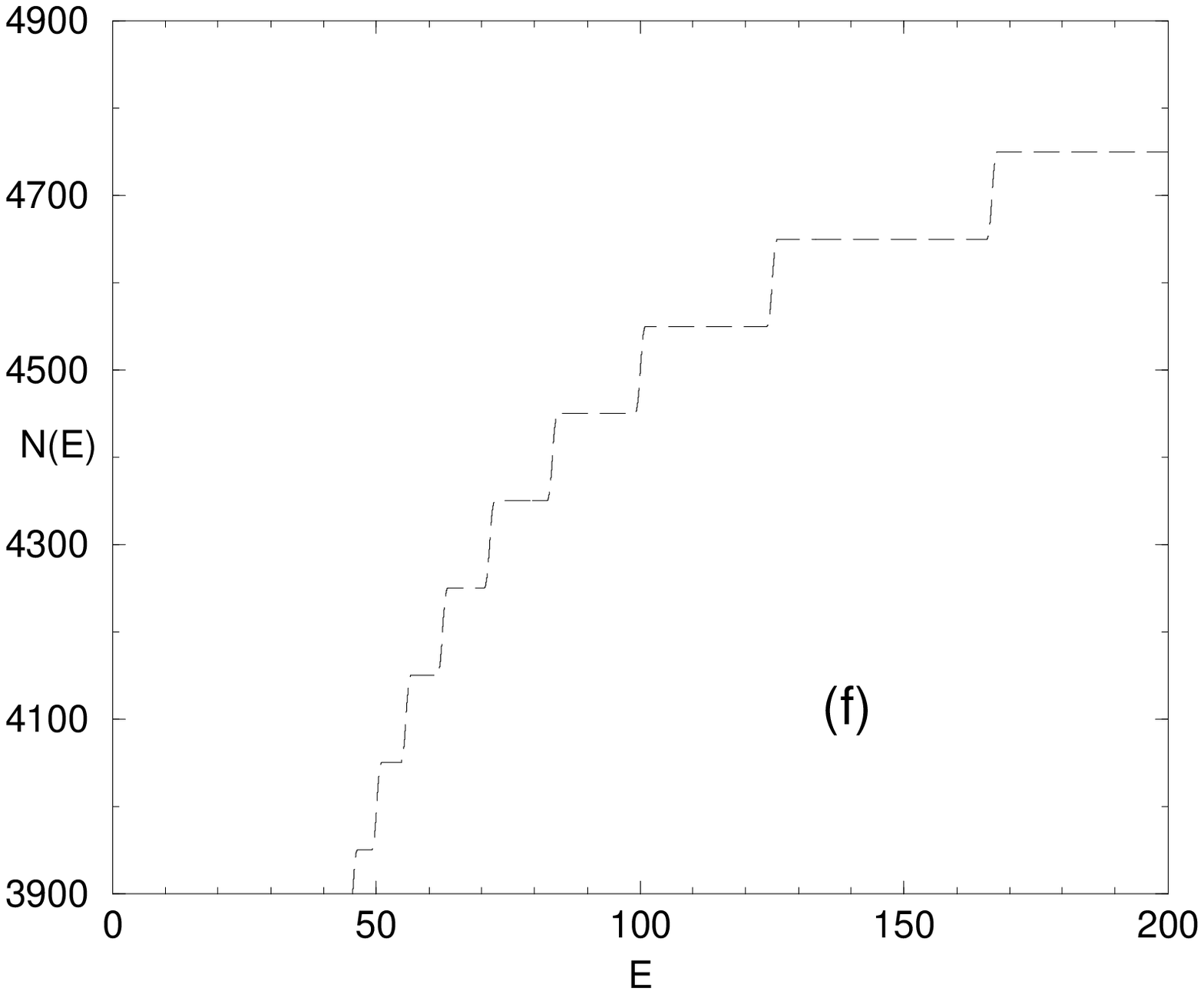}
}
\caption{ Density of states $\rho_2(E)$ for $L=L_1=100$.  
({\bf a}) 2FP states ($U=0$), insert: density around $E \approx 0$ 
(normalized to $1$); ({\bf b}) Coulomb $(U=5)$; 
({\bf c}) Hubbard ($U=35$) The main band of $L(L-1)/2$ states 
remains centered around $E=0$ together with $L$ molecular states 
at $E\approx 2U$.  Coulomb repulsion at ({\bf d}) $U=70$ and 
({\bf e}) $U=500$. Each subband is centered at $E\approx U/d$ 
with $d=0, \ldots, L/2$. ({\bf f}) Integrated density showing that 
each subband has $L$ states for Coulomb repulsion ($U=500$).}
\label{fig1}
\end{figure}

 When one compares the two repulsions, a first 
difference appears in the density of states $\rho_2(E)$.
In the limit $U\rightarrow \infty$, Hubbard repulsion splits 
\cite{tip2} the TIP band in two parts: a small band of $L$ ``molecular 
states'' of high energy $\approx U+2W v_i$ corresponding to two 
electrons localized on the same site $i$ and a main band of 
$L(L-1)/2$ ``hard core boson'' states which remain at the same 
small energies for $U \rightarrow \infty$ and $U \rightarrow 0$.
The ``hard core boson'' states are given by the 
resymmetrization \cite{silvestrov} of Slater determinants corresponding 
to electrons in the one body state $|\alpha>$ and $\beta>$ respectively. Those 
states do not feel on site interaction, are not coupled to one another 
and become decoupled from the molecular states of much larger 
energies when $U \rightarrow \infty$. When one takes rigid BCs, the 
resymmetrization is simple and one has exactly 
$E_{\alpha\beta} (U \rightarrow 0) = E_{\alpha\beta} (U \rightarrow \infty)$ 
for $\alpha \neq \beta$. 
For two electrons in a ring enclosing a flux $\phi$ the resymmetrization 
is more subtle and $E_{\alpha\beta} (\phi+\phi_0/2, 
U \rightarrow 0) = E_{\alpha\beta} (\phi, U \rightarrow \infty)$. 
For periodic BCs, $E_{\alpha\beta}$ ($\alpha \neq \beta$) goes to the 
corresponding 2FP eigenenergy with anti periodic BCs. 
 In contrast to Hubbard repulsion where the majority of the TIP  
energy levels does not feel the interaction when $U \rightarrow \infty$, 
excepted $L$ ``molecular'' states, Coulomb repulsion eventually crystallizes 
all the TIP states as two particle ``molecules''. When $L$ is even, the sizes of 
the molecules are $d=0, \ldots, L/2$ and the spectrum is split in $L/2$ subbands of $L$ 
states ($d\neq L/2$) and one subband of $L/2$ states ($d=L/2$), each of them 
centered around an energy $U/d$. When $N$ is odd, one has $(L+1)/2$ subbbands 
of $L$ states. Without disorder, each subband shrinks onto a single $L$-fold 
degenerate state 
obtained from successive translations of the ``molecules'' by one 
lattice spacing, a degeneracy which is broken by the random potentials. 
The different densities $\rho_2(E)$ induced by the two repulsions 
are illustrated in Fig. \ref{fig1}, for a chain of size $L=L_1=100$ 
and various interaction strengths $U$.

\section{Crossover between two preferential eigenbases}
\label{sec:crossover}

\begin{figure} 
\centerline{ 
\epsfxsize=8cm \epsfysize=7cm
\epsffile{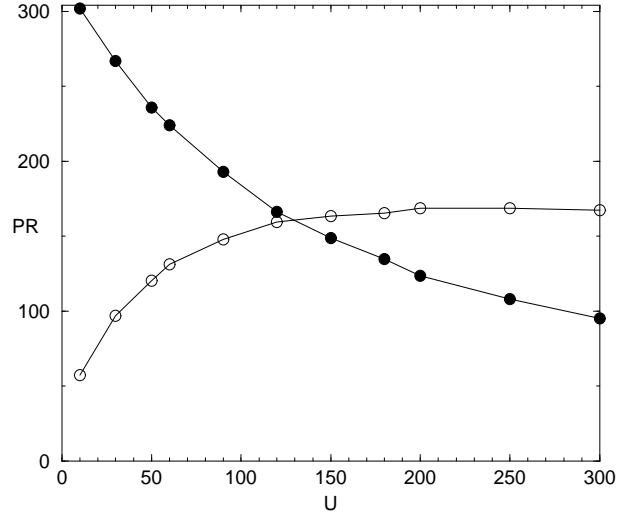} } 
\caption{Participation ratios $PR_{0}$ ($\circ$) and $PR_{\infty}$ 
($\bullet$) of the TIP eigenstates of energies $\approx 2U/L$ 
onto the $U=0$ and $U=\infty$ eigenbases respectively. $L=L_1=50$. 
}
\label{fig9} 
\end{figure}

 When one turns on Hubbard repulsion, it has been detailed in 
Ref. \cite{tip2} how the TIP system goes from the 2FP basis 
towards the ``hard core boson basis'' 
when $U \rightarrow \infty$. The interaction 
threshold $U_c = (24)^{1/4} t/2$  was defined (for energies 
near the band center) as the fixed point of the duality 
transformation mapping the distribution of the interaction matrix 
elements $\propto U/t$ which couple the 2FP states onto the 
distribution of the kinetic energy matrix elements $\propto t/U$ 
which couple the hard core boson states. At $U \approx U_c$, the 
2FP basis ceases to be preferential compared to the hard core boson 
basis. Being unable to extend this duality argument for 
Coulomb interaction, we study the participation ratios 
$PR_{0}$  of the TIP wavefunctions $|\Psi\rangle$ onto the 2FP eigenbasis 
and $PR_{\infty}$ onto the basis built out from symmetrized products of 
site orbitals (site basis). This later basis describes the 
correlated ``molecules'' created when $U \rightarrow \infty$. 
\begin{eqnarray}
PR_{0} & = & (\sum_{\alpha\beta} |\langle \alpha\beta | \Psi \rangle|^4)^{-1} 
\nonumber \\
PR_{\infty} & = & (\sum_{ij} |\langle ij | \Psi \rangle|^4)^{-1}
\end{eqnarray}
This allows us to extend for Coulomb repulsion the concept of a 
crossover threshold $U_c$ where the $U=0$ eigenbasis ceases to be 
preferential compared to the $U=\infty$ eigenbasis. As shown in 
Fig. \ref{fig9}, one has $U_c \approx 120$ when $L=L_1=50$ for 
Coulomb repulsion.

\section{Interaction matrix elements in the 2FP basis}
\label{sec:matrix elements}

 Before discussing the TIP spectral statistics, 
it is useful to study the structure of the interaction matrix elements 
coupling the 2FP states. The 2FP wave function have components 
$\Psi_{\alpha,\beta}(n,m)$ 
on the sites $|nm>$ given by 
$$
<\alpha\beta|nm>= \frac{\psi_{\alpha}(n)\psi_{\beta}(m)+
\psi_{\alpha}(m)\psi_{\beta}(n)}{\sqrt{2}}.
$$
We denote 
\begin{equation}
Q_{\alpha\beta}^{\gamma\delta}(0)= \sum_{n=1}^{L}  
\psi^*_{\alpha}(n)\psi^*_{\beta}(n)\psi_{\gamma}(n)\psi_{\delta}(n)
\end{equation} 
and 
\begin{equation} 
Q_{\alpha\beta}^{\gamma\delta}(p)=\sum_{n=1}^{L}  
\frac{\psi^*_{\alpha}(n)\psi^*_{\beta}(n+p)\psi_{\gamma}(n)\psi_{\delta}(n+p)}
{|p|} + perm 
\nonumber
\end{equation}
where $perm$ means the terms obtained after permuting 
$(\alpha\leftrightarrow\beta), (\gamma\leftrightarrow\delta,)$ 
and $(\alpha\leftrightarrow\beta,\gamma\leftrightarrow\delta)$ 
for $p \neq 0$.

In the 2FP eigenbasis, the interaction matrix elements are 
\begin {equation}
\hspace{-1cm}
\langle\alpha\beta|{\cal U}|\gamma\delta\rangle
= 4U Q_{\alpha\beta}^{\gamma\delta} (0)= U H_{\alpha\beta}^{\gamma\delta}  
\end {equation}
for Hubbard repulsion, 
and
\begin {equation}
\langle\alpha\beta|{\cal U}|\gamma\delta\rangle 
 = U( 4 Q_{\alpha\beta}^{\gamma\delta} (0) + 2 
\sum_{p=1}^{L/2} Q_{\alpha\beta}^{\gamma\delta} (p))
= U C_{\alpha\beta}^{\gamma\delta}
\end {equation}
for Coulomb repulsion.

\begin{figure}
\vspace{1cm}
\centerline{ 
\epsfxsize=6cm \epsfysize=6cm
\epsffile{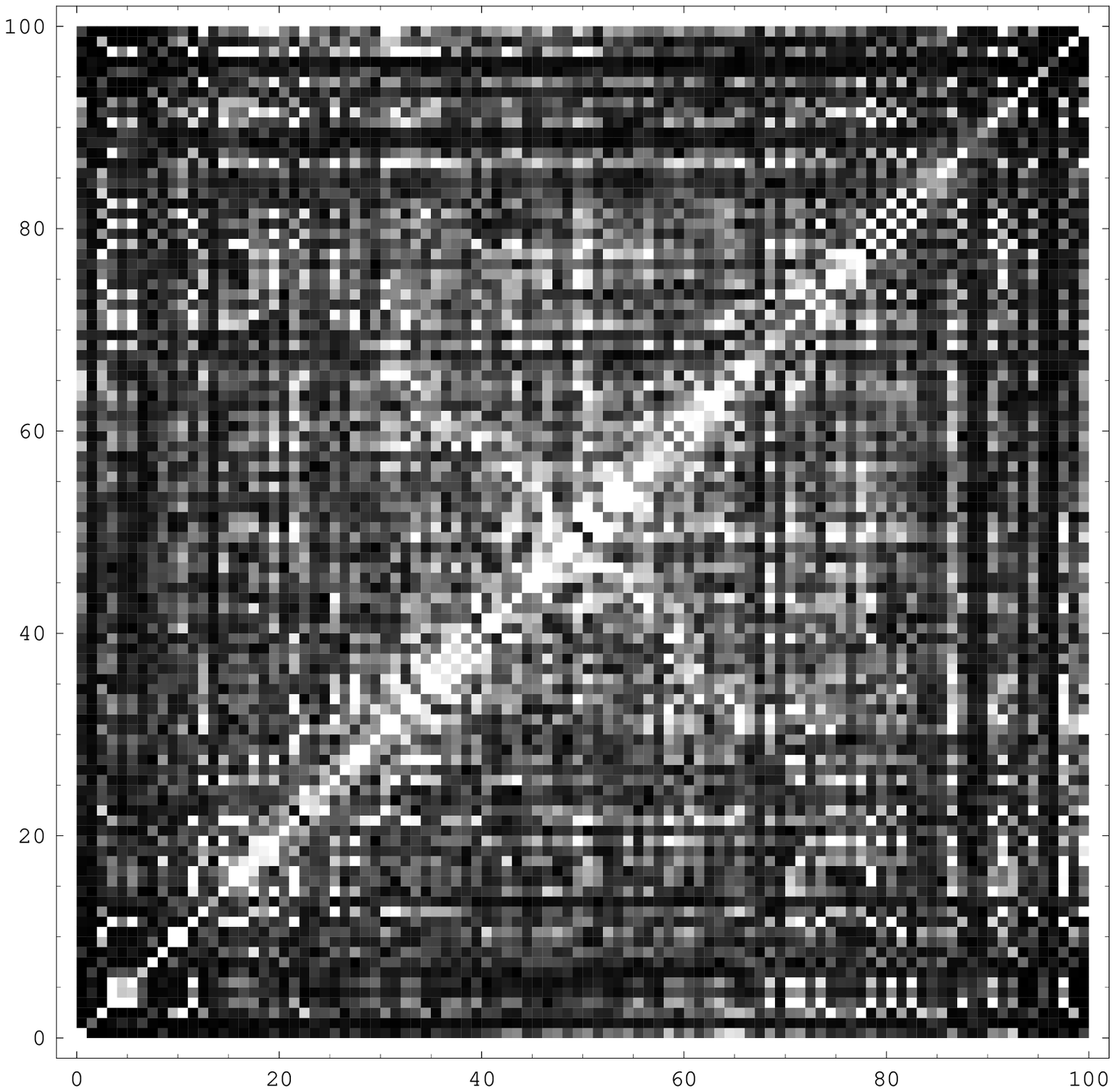}
}
\centerline{  
\epsfxsize=6cm \epsfysize=6cm
\epsffile{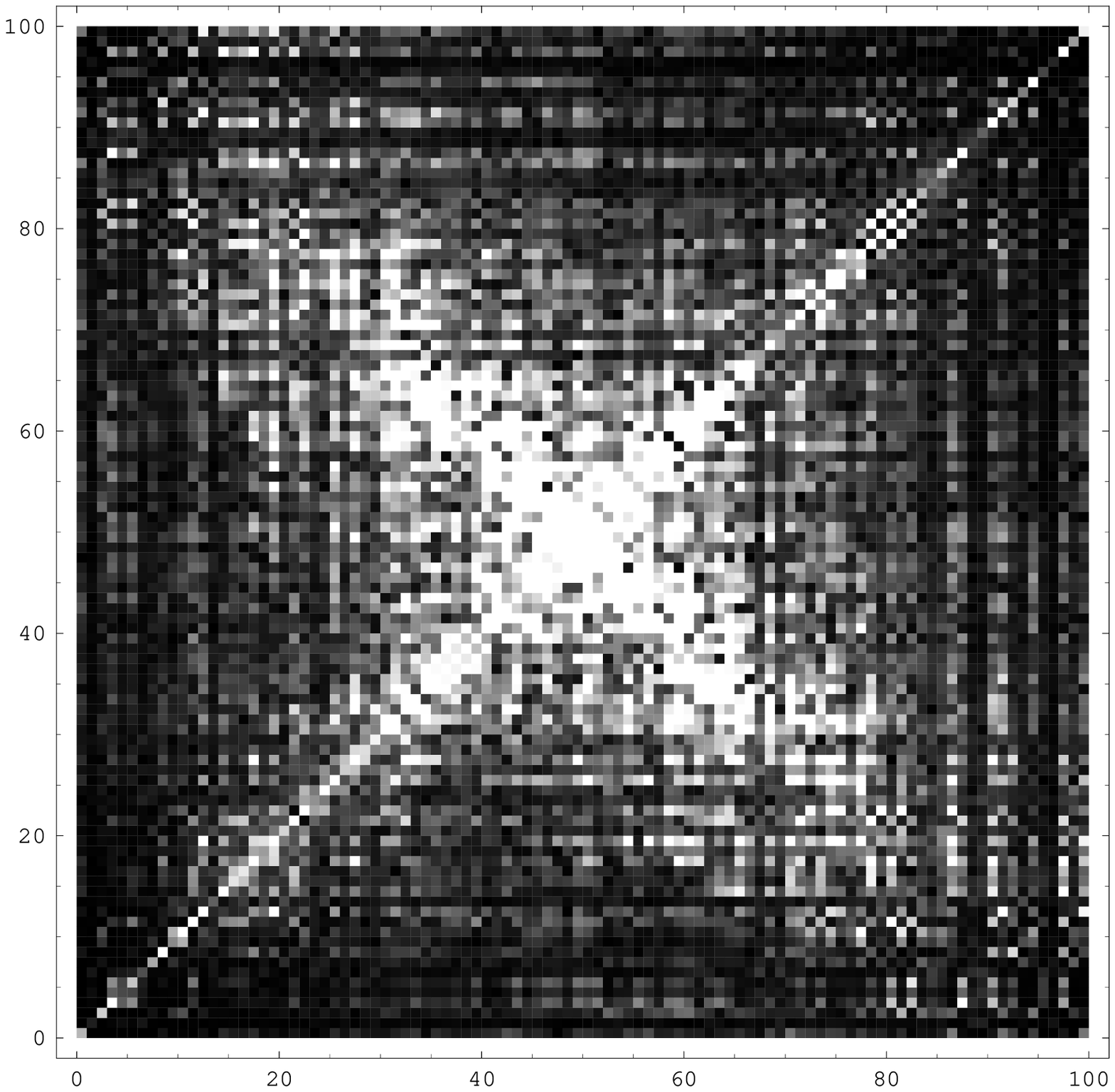}
}
\caption{Magnitude of the absolute value of the interaction matrix 
elements coupling a given state $|\alpha\alpha \rangle$ (with 
$\epsilon_{\alpha} \approx 0$) to $L^2$ states $|\gamma\delta\rangle$ 
for a typical sample with $L=2L_1=100$. The $L^2$ absolute values 
are given in the plan ($\gamma,\delta$). The one particle states are 
ordered by increasing values of the energy. The white (black) 
points correspond to maxima (minima) with a linear scale of graduation. 
Top figure: Hubbard $H_{\alpha\alpha}^{\gamma\delta}$;  
Bottom figure: Coulomb $C_{\alpha\alpha}^{\gamma\delta}$
}
\label{fig2}
\end{figure}
 
In the absence of disorder and with periodic BCs, 
the one body states are plane waves 
$\psi_{\alpha}(n)=(\exp i k_{\alpha}n)/\sqrt{L}$ and the interaction 
matrix elements ${\cal U}_{\alpha\beta}^{\gamma\delta}$ 
only couple 2FP states of same momentum $K=K_{\alpha\beta}=
k_{\alpha}+k_{\beta}=K_{\gamma\delta}$. 
For Hubbard, one has when $L_1 \rightarrow \infty$
\begin{equation}
H_{\alpha\beta}^{\gamma\delta} 
\rightarrow \frac{4}{L} \delta_{K_{\alpha\beta},K_{\gamma\delta}}. 
\label{EqY}
\end{equation}
The interaction matrix has a block diagonal form. If $L$ is odd, 
one has $L$ blocks of size $N_s=(L+1)/2$. If $N$ is even, one has 
$L/2$ blocks of size $N_s=L/2$ and $L/2$ other blocks 
of size $N_s=L/2+1$. The $N_s$ TIP eigenenergies $E_n (K)$ of same 
momentum $K$ are given by the $N_s$ solutions of: 
\begin{equation}
\sum_{\gamma\delta} \frac {1}{E_n(K)-E_{\gamma\delta}}=\frac{L}{4U}
\label{EqVP}
\end{equation} 
where $E_{\gamma\delta}=2 \cos k_{\gamma} + 2 \cos k_{\delta}$.
 The $E_n(K)$ alternate with the 2FP energies $E_{\gamma\delta}$ 
of same momentum. The TIP spectrum is the uncorrelated sum of $L$ such 
series of different momenta $K$.

For Coulomb, the previous block diagonal structure is preserved 
when $L_1 \rightarrow \infty$, but each block becomes more complex:
\begin{eqnarray} 
& &C_{\alpha\beta}^{\gamma\delta}  \rightarrow \frac{4}{L} 
\delta_{K_{\alpha\beta},K_{\gamma\delta}} \times \\
& & (1 +  \sum_{p=1}^{L/2} \frac{\exp-i [(k_{\beta}-k_{\gamma})p]}{2p} 
+ \sum_{p=1}^{L/2} \frac{\exp-i [(k_{\beta}-k_{\delta})p]}{2p} \nonumber \\
& & +\sum_{p=1}^{L/2} \frac{\exp-i [(k_{\alpha}-k_{\gamma})p]}{2p} 
+ \sum_{p=1}^{L/2} \frac{\exp-i [(k_{\alpha}-k_{\delta})p]}{2p} \nonumber)
\label{EqYY}
\end{eqnarray}

\begin{figure} 
\centerline{ \epsfxsize=8cm \epsfysize=6cm
\epsffile{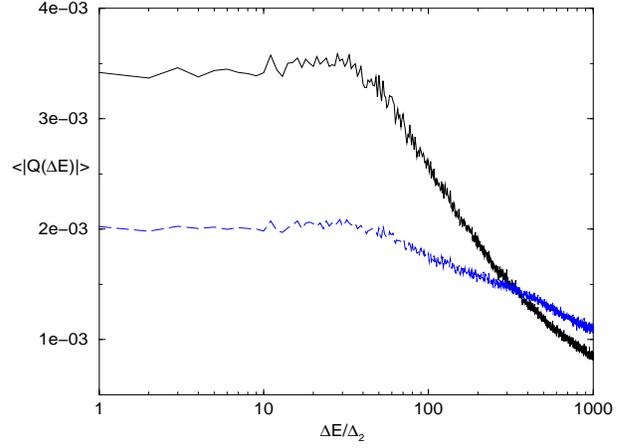} } 
\caption
{ Disorder average amplitudes 
$<|H_{\alpha\alpha}^{\gamma\delta}|>$ (dashed line) and  
$<|C_{\alpha\alpha}^{\gamma\delta}|>$ (continuous line) of 
the matrix elements coupling 2FP states separated 
by an energy $\Delta E = |E_{\alpha\alpha}-E_{\gamma\delta}|$ 
(in units of $\Delta_2$) for $L=2L_1=100$.
}  
\label{fig3} 
\end{figure}
  
 In the presence of a random potential, the absolute values of the matrix 
elements can be given using a linearly graduated grey scale in the plane 
$(\gamma,\delta)$ for a given 2FP state $|\alpha\beta>$, the one body 
states $|\gamma>$ and $|\delta>$ being ordered by increasing energies. 
When $W$ is sufficiently small, TIP momentum remains almost conserved and 
one can see in the plane $(\gamma,\delta)$ a white cross made by the 
two diagonals if $\alpha=\beta$. The first diagonal corresponds to 
coupling  to other states $|\gamma\delta>$ with $|\gamma> 
\approx |\delta>$, the second to coupling to 2FP states 
$|\gamma\delta>$ close in energy ($E_{\alpha\beta} \approx E_{\gamma\delta}$). 
When $W$ is larger, Coulomb and Hubbard give rise to a different pattern 
in the $(\gamma\delta)$ plane: energy-momentum conservation remains 
partially preserved by Coulomb repulsion (the second diagonal persists) 
and is lost by Hubbard repulsion, as shown in Fig. \ref{fig2} 
for $L\approx2L_1=100$. 2FP states $|\alpha\alpha>$ 
with $\epsilon_{\alpha} \approx 0$ are considered in Fig. \ref{fig2}, 
but similar conclusions can be drawn from arbitrary 2FP states 
$|\alpha\beta>$. 

 To explain why $C_{\alpha\beta}^{\gamma\delta}$ continues to mainly 
couple $|\alpha\beta>$ to states $|\gamma\delta>$ nearby in energy, 
we note that when $L_1$ is  finite, disorder smears the sharp 
delta function into a broader gaussian peak of width $\sigma \propto 
L_1^{-1}$. $\langle \delta_{K_{\alpha\beta},K_{\gamma\delta}} \rangle 
\rightarrow \exp -(K_{\alpha\beta}-K_{\gamma\delta})^2 / (2 \sigma^2)$ and 
$\langle \exp -i(k_{\alpha}-k_{\beta})p \rangle \rightarrow $\\$ 
\exp -(k_{\alpha}-k_{\beta})^2 / (2 \sigma^2)$, one gets respectively
$$
H_{\alpha\beta}^{\gamma\delta} \propto
\exp-\frac {(K_{\alpha\beta}-K_{\gamma\delta})^2}{2 \sigma^2}
$$
and 
\begin{eqnarray}
\label{EqC}
\hspace{-0.5cm}
&C_{\alpha\beta}^{\gamma\delta} & \propto   
\exp (-\frac{(K_{\alpha\beta}-K_{\gamma\delta})^2}{2 \sigma^2}) \times \\ 
& &[2 - \ln (\frac{|k_{\beta}-k_{\gamma}|}{\sigma} - \ln (\frac{|k_{\beta}-k_{\delta}|}{\sigma})\nonumber\\
& &  - \ln (\frac{|k_{\alpha}-k_{\gamma}|}{\sigma}) - \ln (\frac{|k_{\alpha}-k_{\delta}|}{\sigma})]\nonumber
\label{EqX}
\end{eqnarray}

 This explains the behavior shown in Fig. \ref{fig3} where the disorder 
averaged amplitude of the hopping terms are given as a function of the energy 
difference. The $L$ hopping terms between the 2FP states $|\gamma\delta>$ nearby in energy 
$(\Delta E \leq\Delta_2 L/2)$ are much smaller for Hubbard than for 
Coulomb. For larger energy separation $\Delta E$, 
there is a $\ln (|\Delta E/\Delta_2|)$ 
decay  which is more pronounced for Coulomb than for Hubbard.

\section{TIP diffusion and 2FP lifetime}
\label{sec:lifetime}

 Coulomb repulsion couples a density $\approx \rho_2$  
of 2FP states nearby in energy. Hubbard repulsion effectively couples a 
smaller density $\rho_2^{eff} < \rho_2$, as explained in Ref. \cite{tip1}. 
Let us review three consequences of this difference.

(i) TIP diffusion: In the first studies \cite{shepelyansky,imry,fmgp} 
of the TIP problem, a density $\rho_2(L_1)$ of 2FP states coupled by 
the interaction was assumed for $L \approx L_1$. Under this assumption, it 
was predicted that the TIP dynamics should exhibit interaction assisted 
diffusion on scales $L_1 < L < L_2$, the time evolution of the TIP center 
of mass $R_2$ being given \cite{fmgp} by: 
\begin{equation}
R_2(t) \approx \sqrt{D_2 (t) t}
\end{equation}
with $D_2(t)$ is roughly constant, up to $log (t)$ corrections \cite{fmgp}. 
In Ref. \cite{tip3}, a much slower propagation $R_2 \propto \log t$ 
was observed for Hubbard repulsion, attributed to the weak density 
$\rho_2^{eff}(L_1)$ of effectively coupled states. One expects that 
the original prediction will be at least partially restored for 
Coulomb repulsion. 

(ii) TIP localization: The interaction assisted propagation stops  
at a scale $L_2$ characteristic of TIP localization. 
Assuming $ |<\alpha\beta |{\cal U}|\gamma\delta>|^2 \propto U^2/L_1^3$ for 
$L\approx L_1$, the enhancement factor $L_2/L_1$ was originally 
given by the estimate:  
\begin{equation}
\frac{L_2}{L_1} \propto |<\alpha\beta|{\cal U}|\gamma\delta>|^2 \rho_2 (L_1)
\propto L_1
\end{equation}
It was pointed out in Ref. \cite{tip1} that, since $\rho_2(L_1) 
\propto L_1^{2} \rightarrow \rho_2^{eff} (L_1) \propto L_1^{1.75}$ 
in the presence of Hubbard repulsion, the enhancement factor should be 
weaker ($L_2/L_1 \propto \sqrt{L_1}$), a prediction confirmed by 
numerical calculations. It is likely that $L_2/L_1 \propto L_1$ will be 
a better estimate for Coulomb repulsion.

(iii) 2FP lifetime: The inverse lifetime $(\Gamma_{\alpha\beta})$ 
of a 2FP state $|\alpha\beta\rangle$ is given by Fermi Golden rule: 
\begin{equation}
\Gamma_{\alpha\beta} \propto \sum_{\gamma\delta} 
|<\alpha\beta |{\cal U}|\gamma\delta>|^2 
\delta (E_{\alpha\beta}-E_{\gamma\delta})
\label{Eq.FGR}
\end{equation}
If the density of coupled states is $\rho_2(L_1) \approx L_1^{2}$, 
a reasonable estimate for Coulomb repulsion, one gets
$$
\Gamma_{\alpha\beta} (L=L_1) \propto  
|<\alpha\beta |{\cal U}|\gamma\delta>|^2 \rho_2(L_1) \propto L_1^{-1}.
$$
 In contrast, the lifetime 
should be longer for Hubbard repulsion: 
$$
\Gamma_{\alpha\beta} (L=L_1) \propto  
|<\alpha\beta |{\cal U}|\gamma\delta>|^2 \rho_2^{eff}(L_1) \propto L_1^{-1.65}
$$  
since the density \cite{tip1} of coupled 
states (by the second moment of $|<\alpha\beta |{\cal U}|\gamma\delta>|$) is 
$\rho_2^{eff}\propto L_1^{1.35}$ for a 2FP state $|\alpha\alpha>$.  

\begin{figure}
\vspace{-0.5cm}
\centerline{ \epsfxsize=8cm \epsfysize=7cm
\epsffile{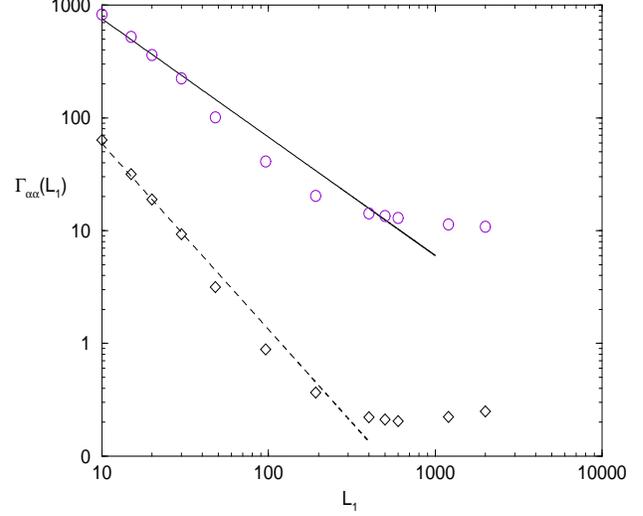} } 
\caption{Inverse lifetime $\Gamma_{\alpha\alpha} (L_1)$ of a 2FP state 
$|\alpha\alpha>$ with $\epsilon_{\alpha} \approx 0$ for Hubbard ($\diamond$) 
and Coulomb ($\circ$) repulsion respectively. $L=420$. The continuous 
(dashed) line corresponds to $\Gamma_{\alpha\alpha} \propto L_1^{-1.65}$ 
($\propto L_1^{-1.05}$). 
}

\label{fig4} 
\end{figure}

 We have checked this prediction. In Fig. \ref{fig4}, the inverse lifetime 
calculated in a chain of length $L=420$ is shown as a function of $L_1$. When 
$L_1<L$,  one has 
$\Gamma_{\alpha\alpha} \propto L_1^{-1.65}$ for Hubbard repulsion 
while $\Gamma_{\alpha\alpha} \propto L_1^{-1.05}$ for Coulomb repulsion, 
close to a simple decay $\propto L_1^{-1}$. When 
$L$ becomes larger than $L_1$, the lifetime becomes $L$-independent 
for Hubbard, and may continue to weakly decay as a function of $L$ 
for Coulomb. Fig. \ref{fig4} shows us that this possible decay remains 
negligible. 

\section{Two particle spectral statistics}
\label{sec:statistics}

 We now study how the spectral statistics depend on the interaction 
strength $U$ for the two repulsions, using the distribution 
$P(s)$ of energy spacings between consecutive levels and the variance 
$\Sigma_2(E)$ of the number of levels inside an energy window of width E. 
We consider energy levels in the bulk of the low energy sub-band: 
$E\approx 0$ for Hubbard repulsion and $E\approx 2U/L$ for Coulomb 
repulsion (see Fig. \ref{fig1}). We take $L=L_1$ for having the largest 
interaction matrix elements, and hence the maximum mixing of the 2FP states. 

After unfolding the spectra, one expects 
\begin{equation}
P(s) \approx P_W(s)\approx \frac{\pi s}{2} \exp(-\frac{\pi}{4}s^2)
\end{equation}
and 
\begin{equation}
\Sigma_2(E) \approx \Sigma_2^W(E) \approx 
\frac{2}{\pi^2}(ln(2\pi E)+\gamma +1-\frac{\pi^2}{8})
\end{equation}
($\gamma$ being the Euler constant) for correlated levels having 
Wigner-Dyson statistics, whereas one should have Poisson statistics 
with 
\begin{equation}
P_P(s)=\exp(-s)
\end{equation}
and
\begin{equation}
\Sigma_2^P(E)=E
\end{equation}
for uncorrelated levels. 

\begin{figure}
\vspace{-0.25cm} 
\centerline{ \epsfxsize=8cm \epsfysize=7cm
\epsffile{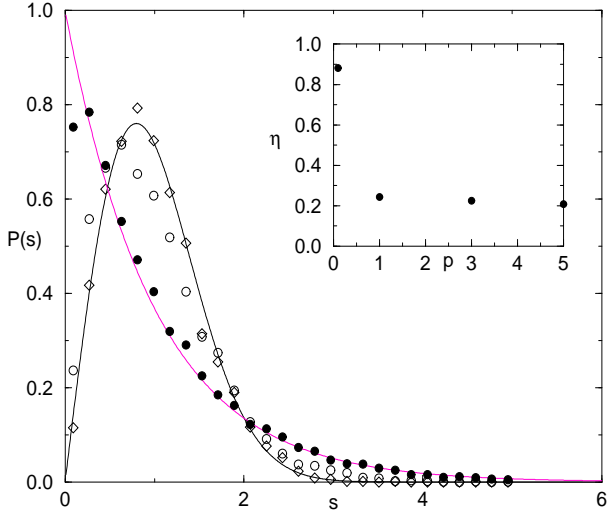} } 
\caption{Spacing distribution P(s) for Hubbard ($\bullet$) 
and Coulomb ($\diamond$) repulsions at $U=70$ and $L=L_1=50$. 
The symbols $(\circ)$ correspond to a medium range interaction 
truncated to $p = 1$. The levels are taken around $E \approx 2U/L$ 
for Coulomb repulsion and around $E \approx 0$ otherwise. The continuous 
lines correspond to $P_P(s)$ and $P_W(s)$ respectively. 
Insert: Parameter $\eta$ around $E \approx 0$ 
as a function of the range $p$ of the interaction.}
\label{fig5} 
\end{figure}

\begin{figure} 
\centerline{ \epsfxsize=8cm \epsfysize=7cm
\epsffile{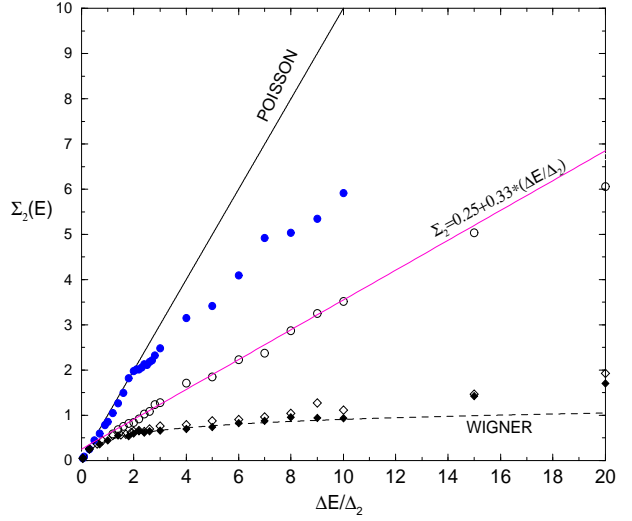} } 
\caption{
Number variance $\Sigma_2$ as a function of the energy interval 
$E$ in units of TIP level spacing $\Delta_2$ (same symbols as 
in Fig. \ref{fig5} excepted the filled diamonds (Coulomb at 
$U=100$ and $L=L_1=100$). 
}
\label{fig6} 
\end{figure}

 The TIP spectra are well described by Wigner-Dyson statistics for an 
intermediate Coulomb repulsion. In Fig. \ref{fig5} and in Fig. \ref{fig6}, 
one can see indeed that the level spacing distribution and the number 
variance $\Sigma_2$ are well described by the Wigner surmise $P_W(s)$ and 
$\Sigma_2^W(E)$ respectively near the intermediate interaction strength 
$U_c \approx 120$ for which the TIP system does not have a preferential 
eigenbasis (see Fig. \ref{fig9}). As shown by $\Sigma_2(E)$ (Fig. \ref{fig6}),
Wigner-Dyson spectral rigidity is established over an energy interval 
containing a few levels.  In contrast, Hubbard repulsion 
can only yield \cite{tip2} at the corresponding $U_c$ ($\approx 1$)
\begin{equation}
P_{SP} \approx 4s \exp-(2s)
\label{EqSP}
\end{equation}
and 
\begin{equation}
\Sigma_2 \approx 0.16 + 0.41 E
\label{EqNV}
\end{equation}
respectively.

To study how the spacing distribution depends on $U$, we use the spectral 
parameter $\eta$ defined by 
\begin {equation}
\eta(P,U)=\frac{\int_0^b ds[P(s)-P_W(s)]}{\int_0^b ds[P_P(s)-P_W(s)]}
\end {equation}
with $b=0.4729$. 
For $U=0$, the consecutive levels are essentially uncorrelated and 
$\eta \approx 1$. The curves $\eta(U)$ given in Fig. \ref{fig7} 
show us a striking difference between the spectral statistics 
yielded by the two repulsions. 

\begin{figure} \centerline{ \epsfxsize=8cm \epsfysize=7cm
\epsffile{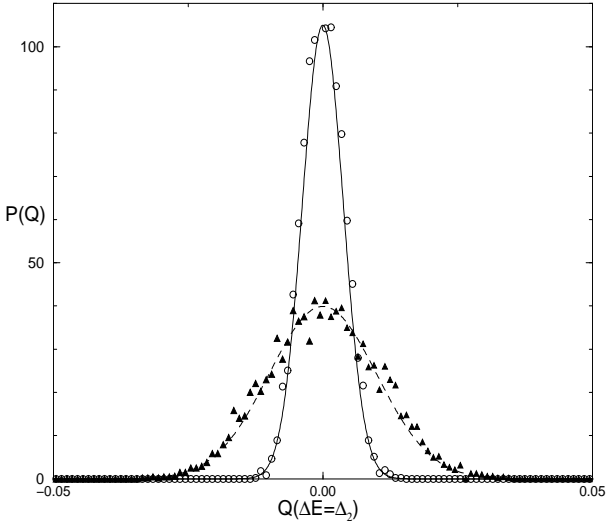} } 
\caption{Distribution $P(Q)$ of the interaction matrix elements 
$Q= H_{\alpha\beta}^{\gamma\delta}$ (Hubbard: circle) or 
$Q=C_{\alpha\beta}^{\gamma\delta}$ (Coulomb: triangle) coupling 
nearest neighbor 2FP states (energy separation $\approx \Delta_2$) 
for $L=L_1$. The distributions are fitted with gaussian curves of 
variance $\sigma^2 =0.01^2$ (Coulomb, dashed line) and $\sigma^2=0.0038^2$ 
(Hubbard, continuous line).}
\label{fig8} 
\end{figure}

\begin{figure}
\centerline{ \epsfxsize= 8cm \epsfysize=7cm
\epsffile{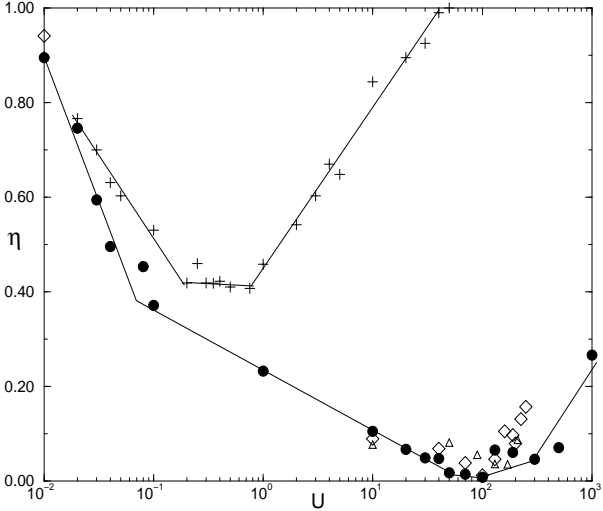} } 
\caption{ Spectral parameter $\eta$ as a function of $U$ for 
$L=L_1$. Levels around $E\approx 2U/L$ for Coulomb repulsion 
with $L= 100 (\bullet)$, $50 (\diamond)$ and $65(\triangle)$. 
Levels around $E \approx 0$ for Hubbard repulsion $(+)$
with $L=100$. 
}
\label{fig7} 
\end{figure}

 For $L=L_1=100$, let us estimate the interaction threshold $U^*$ 
for which the interaction matrix elements coupling consecutive 2FP 
states becomes of the order of their energy separation $\Delta_2$. 
$\rho_2 \approx 0.25$ (density normalized to $1$) gives 
$\Delta_2 \approx (\rho_2 L(L+1)/2)^{-1} \approx 4/5050$ 
(see insert of Fig. \ref{fig1}) for the energy spacing between 
consecutive 2FP levels around $E=0$.  The off-diagonal interaction matrix 
elements $Q= H_{\alpha\beta}^{\gamma\delta}$ (Hubbard) or 
$Q=C_{\alpha\beta}^{\gamma\delta}$ (Coulomb) coupling 
consecutive 2FP levels are normally distributed, as 
shown in Fig. \ref{fig8}. The root mean square of $Q$ is small for 
Hubbard ($\approx 0.0038$) and larger for Coulomb 
($\approx 0.01$). This gives $U^*_C \approx 0.08$ for 
Coulomb and $U^*_H \approx 0.2$ for Hubbard. 

 When $U$ increases but remains lower than $U^*$, there is first a perturbative regime discussed in 
Refs. \cite{wp,wpi} where the interaction yields Rabi oscillations 
between consecutive 2FP states at a frequency given by the absolute 
value of the coupling interaction matrix element. Moreover, the energy 
range $E_U$ under which one has level repulsion is given \cite{wp} by 
this Rabi frequency. When $U$ becomes equal to $U^*$, 
one has a transition from this perturbative regime towards an effective 
Fermi golden rule decay of the 2FP states and the characteristic 
range $E_U$ over which Wigner-Dyson rigidity occurs becomes \cite{wp,wpi}
proportional to the square of the amplitude of the coupling matrix 
elements: $E_U \propto U$ when $U < U^*$ and $E_U \propto U^2$ when 
$U > U^*$. $U=U^*$ is the interaction threshold where the spectral 
statistics are intermediate between Poisson and Wigner ($\eta \approx 
0.39$ when $P(s)\approx P_{SP}(s)$). Looking at Fig. \ref{fig7}, 
one can see that $\eta$ decreases as a function of $U$ down to 
the characteristic value $\eta^* \approx 0.39 $ reached 
when $U \approx U^*_C \approx 0.08$ for Coulomb repulsion, and when 
$U \approx U^*_H \approx 0.2$ for Hubbard repulsion. 

 For Coulomb repulsion, $\eta$ continues to decrease when $U \geq U^*_C$ 
down to $\eta=0$ with the slower $U$ dependence characteristic of the  
effective golden rule decay of the 2FP states. The Wigner-Dyson distribution 
$P_W(s)$ is fully established at $U\approx U_c$, i.e. when the TIP system 
is exactly as far from the $U=0$ eigenbasis than from the $U=\infty$ 
eigenbasis.  When $U \geq U_c$, integrability is slowly restored and 
$\eta \rightarrow 1$ as $U \rightarrow \infty$. 

 For Hubbard repulsion, the spectral rigidity does not continue to 
increase above $U_H^*$, but saturates to the intermediate critical rigidity 
characterized by Eq. \ref{EqSP} and Eq. \ref{EqNV} for $P(s)$ and 
$\Sigma_2(E)$ respectively. Above the fixed point $U_c \approx 1$ of the 
duality transformation, the TIP system becomes 
closer to the $U=\infty$ eigenbasis and the levels become statistically 
uncorrelated. This critical Hubbard regime is a complicated issue where 
the multifractal character of the interaction matrix described in 
ref. \cite{tip1} is very likely relevant. However, one can do the 
following remark. When $L_1 \rightarrow \infty$, the interaction matrix 
is block diagonal, a block corresponding to a pair momentum $K$.
For Hubbard, the $N_s$ TIP level $E_n(K)$ of momentum $K$ are located 
near the 2FP levels of same momentum when $4U/L \rightarrow 0$ or 
$4U/L \rightarrow \infty$. One can assume that they should be near the 
middle of consecutive 2FP levels of same $K$ for $U \approx U_c$. The 
distribution $P(s)$ 
of levels located in the middle of levels with spacing distribution $P_P(s)$ 
is the semi-Poisson distribution $P_{SP}(s)$. If the sequence of 2FP levels 
of momentum $K$ were randomly distributed, the spacing distribution 
of the $N_s$ TIP levels should be given by $P_{SP}(s)$ without disorder. 
On the contrary, the interaction matrix being more random for Coulomb, 
one can expect that the $L$ series of $N_s$ levels of momentum $K$ will 
be driven towards Wigner-Dyson statistics as $U$ increases.
However, though the argument may give hints for the existence of the 
Hubbard $P_{SP}(S)$, it does not explains the behavior of $\Sigma_2 (E)$. 
The breakdown of momentum conservation by the disorder, and the associated 
mixing of the $L$ independent series of $N_s$ levels 
characterizing the clean limit plays a complex role. 

The insert of Fig. \ref{fig5} shows how the spectrum becomes more rigid 
at $U=70$ when the range $p$ of the interaction is increased. 
$\Sigma_2(E)$ displays a similar information in Fig. \ref{fig6}. 

\section{Quantum melting for intermediate Coulomb repulsions}
\label{sec:delocalization}

\begin{figure}
\vspace{-0.5cm}
\centerline{ \epsfxsize= 8cm \epsfysize=7cm
\epsffile{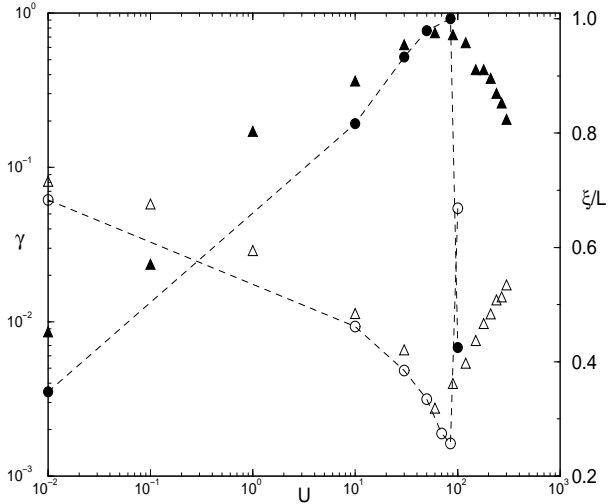} } 
\caption{Crystallization parameter $\gamma$ 
for $L=L_1=50 (\triangle)$ and $100 (\circ)$ and 
the corresponding numbers of occupied sites 
$\xi$ (filled symbols) for Coulomb repulsion 
for TIP levels around $E=2U/L$.}
\label{fig10} 
\end{figure}

 The intermediate Wigner-Dyson regime yielded by Coul- omb repulsion 
corresponds, inside a scale $L_1$, to a complete melting of the localized 2FP 
states previous to crystallization. To show this, we introduce 
two parameters $\gamma$ and $\xi$. For a TIP wavefunction 
$|\Psi\rangle$, we calculate the density $\rho_i = \langle \Psi |c_i^{\dagger}
 c_i | \Psi \rangle$ at site $i$ and the density density correlation 
function $C(r)= (1/2) \sum_i \rho_i \rho_{i-r}$. The participation ratio 
$\xi$ (i. e. the number of sites occupied by a TIP state) is given by 
$2C(0)^{-1}$. The crystallization parameter $\gamma$ is given by the 
difference  $Max C(r) - Min C(r)$, where all translations $r$ 
(including $r=0$) are considered. If the electron density is homogeneous 
(as for an extended liquid state) $\gamma \approx 0$ whereas $\gamma \approx 1$ 
if the two charges are mainly 
located on two different sites, a situation occuring when $U \rightarrow 
\infty$ (Coulomb ``molecule'') and when $U\rightarrow 0$ and $L_1 
\rightarrow 0$ (two electrons located in two minima of the potential). 
The variations of $\xi$ and $\gamma$ are given in Fig. \ref{fig10} when 
Coulomb repulsion $U$ increases. The curves $\gamma(U)$ and $\xi(U)$ are 
correlated to the curve $\eta(U)$ shown in Fig. \ref{fig7}. The curve 
$\xi(U)$ shows us that the TIP states occupy a fraction of the chain without 
repulsion before being uniformly spread over a scale $L_1$ at interaction 
strength $U\approx U_c$ for which there is a crossing of the two curves 
$PR(U)$ given in Fig. \ref{fig9}. For $L=L_1$, one has an interaction 
induced quantum melting of the non interacting glass, yielding quantum 
ergodicity with `` more liquid'' and extended wavefunctions and Wigner-Dyson 
spectral statistics.
 
\section{Conclusions}
\label{sec:conclusions}

 We have studied one of the simplest problems where quantum 
localization and two body interaction are in competition. 
We have seen that the range of the interaction makes important 
differences. One of them is revealed by the study of the spectral 
statistics, providing an intriguing puzzle for quantum 
ergodicity: the Wigner-Dyson statistics shrinking to intermediate 
statistics when the two body repulsion becomes local. 
However, the generic behavior of the TIP system can be summarized 
by three regimes, independently of the interaction range. There is 
the free particle limit dominated by quantum localization, the 
Coulomb limit dominated by the pinning of a correlated system 
of charges (``Coulomb molecule, Wigner crystal'') or by quantum 
localization again (Hubbard repulsion). Between these two limits, 
there is an intermediate regime where one has a maximum mixing of the 
one body states, making the states more extended and the spectrum more 
rigid. In one dimension, this yields a partial delocalization effect 
($L_2$ could be large but remains finite). 
Similar conclusions have been reached from a study of the many body 
ground state \cite{sjwp} of one dimensional spinless fermions at half 
filling. In two dimensions, there are experimental evidences 
\cite{kravchenko} that localization may disappear for intermediate 
Coulomb energy to Fermi energy ratios $r_s$. A new quantum regime has 
been observed \cite{bwp} 
for the ground state of two dimensional spinless fermions at intermediate 
factors $r_s$ where the metallic phase is observed. The nature of the states, 
when the system is far from the free particle (Fermi glass) or the Coulomb 
(Wigner crystal) bases, and the associated transport mechanism, remain to be 
understood. This simple one dimensional study draws our attention to the 
important role played by the range of electron-electron repulsions. 
To know how is or is not screened Coulomb repulsion in low density electron 
systems is then an important issue.

\end{document}